\documentclass[12pt,a4paper]{article}
\usepackage{epsfig}
\begin{document}
\title{Leptonic $Z$ decays in the littlest $Higgs$ model \\with T-parity}
\author{Chong-Xing Yue, Jin-Yan Liu, Shi-Hai Zhu \\
{\small Department of Physics, Liaoning  Normal University, Dalian
116029, P. R. China}
\thanks{E-mail:cxyue@lnnu.edu.cn}}
\date{\today}

\maketitle
\begin{abstract}

\vspace{1cm}

The littlest $Higgs$ model with T-parity (called the $LHT$ model)
predicts the existence of the T-odd leptons, which can generate
contributions to some leptonic processes at the one-loop level. We
calculate their contributions to the leptonic $Z$ decay processes
$Z\rightarrow l\bar{l'}$, $Z\rightarrow l\bar{l}$, and $Z\rightarrow
\nu\bar{\nu}$. We find that the T-odd leptons can give significant
contributions to the branching ratios of these decay processes in
most of the parameter space.  The experimental measurement values
might generate constraints on the free parameters of the $LHT$
model.

 \vspace{2.0cm} \noindent

\end{abstract}
\newpage
\noindent{\bf 1. Introduction}

A precise measurement of gauge boson production for $pp$ scattering
will be crucial at the $LHC$ [1]. At the $LHC$, the gauge boson $Z$
can be copiously produced via the $Drell-Yan$ process and can be
detected through its leptonic $Z$ decay modes. Thus, considering the
contributions of the new physics to the leptonic $Z$ decays is very
interesting, which will be important in testing the standard model
($SM$) and uncovering the possible signals of new physics.

In the $SM$, the lepton flavor-conserving ($LFC$) $Z$ decays
$Z\rightarrow l^{+}l^{-}$ ($l$=$e$, $\mu$, and $\tau$) can proceed
at the tree level. Including $QED$ contributions, the $SM$
prediction values for the branching ratios ($BRs$) of these decay
processes are [2]
\begin{equation}
BR(Z \rightarrow e^{+}e^{-})=3.3346\%,\hspace{1cm} BR(Z \rightarrow
\mu^{+}\mu^{-})=3.3346\% ,
\end{equation}
\begin{equation}
BR(Z \rightarrow \tau^{+}\tau^{-})=3.3338\%,
\end{equation}
and their experimental measurement values are

\begin{equation}
BR(Z \rightarrow e^{+}e^{-})=3.363\pm0.004\% ,\hspace{1cm}
BR(Z\rightarrow\mu^{+}\mu^{-})=3.366\pm0.007\% ,
\end{equation}
\begin{equation}
BR(Z \rightarrow \tau^{+}\tau^{-})=3.370\pm0.0023\%.
\end{equation}
It is obvious that the discrepancy between the experimental and the
$SM$ prediction values is of the order of $1.0$\%. If new physics
models have contributions to the $LFC$ $Z$ decays, this discrepancy
might give constraints on the free parameters of the new physics
models [3].

Since the lepton flavor is conserved in the $SM$, the lepton flavor
violation ($LFV$) decay processes $Z\rightarrow l\bar{l'}$ exist at
least in the one-loop level, and therefore their $BRs$  are
extremely small[4,5]. Their values are far below the experimental
limits obtained at $LEP1$ [2]
\begin{equation}
BR(Z \rightarrow \tau^{\pm}\mu^{\mp})<1.2\times 10^{-5},
\hspace{1cm} BR(Z\rightarrow\tau^{\pm}e^{\mp})<9.8\times10^{-6},
\end{equation}

\begin{equation}
BR(Z \rightarrow \mu^{\pm}e^{\mp})<1.7\times 10^{-6},
\end{equation}
and with the improved sensitivities at $Giga-Z$ [6,7], these numbers
could be pulled down to
\begin{equation}
BR(Z \rightarrow \tau^{\pm}\mu^{\mp})<f\times1.2\times 10^{-8},
\hspace{1cm}
BR(Z\rightarrow\tau^{\pm}e^{\mp})<f\times6.5\times10^{-8},
\end{equation}
\begin{equation}
BR(Z \rightarrow \mu^{\pm}e^{\mp})<2\times 10^{-9}
\end{equation}
with $f=0.2--1.0$. It is very interesting to study the new physics
contributions to the $LFV$ decay processes $Z\rightarrow l\bar{l'}$.
This fact has led to a lot of works related to these decays in the
literature [8,9,10].

In the $SM$, the decay width of the gauge boson $Z$ into each family
neutrino is calculated to be $\Gamma_{\nu\bar{\nu}}=166.3\pm1.5MeV$,
and the current experimental value for the invisible $Z$ decay width
is $\Gamma^{exp}_{inv}=499\pm1.5MeV$ [2]. It is well known that, the
mixing of the active neutrino with the sterile neutrino, additional
generation fermions, or other new weakly interacting particles might
give contributions to the invisible $Z$ decay width. Using the
experimental value of the invisible $Z$ decay width, one can obtain
constraints on the new physics [11,12,13].

The leptonic $Z$ decays are free from the long distance effects and
thus are clean. On the other hand, they carry a considerable
information about the free parameters of the model used. Therefore,
it is worthwhile to analyze these decay processes in the context of
the new physics models. In the present work, we first consider the
$LFV$ coupling vertex $Zl\bar{l'}$ induced by the new particles in
the framework of the littlest $Higgs$ model with T-parity (called
the $LHT$ model) [14] and calculate the branching ratio $BR(Z
\rightarrow l\bar{l'})$. Then we study the contributions of the
$LHT$ model to the $LFC$ decay process $Z\rightarrow l\bar{l}$ and
analyze whether the $LHT$ effect overcomes the discrepancy of the
$BR's$ value between the experimental and the $SM$ prediction
results. Finally, in the context of the $LHT$ model, we calculate
the invisible $Z$ decay width $\Gamma_{inv}$ and compare our
numerical results with the experimental values for $\Gamma_{inv}$.

The layout of the present paper is as follows: After giving a brief
of review the essential features of the $LHT$ model, we study the
branching ratios of the $LFV$ decay process $Z \rightarrow
l\bar{l'}$ in section 2. The contributions of the new particles
predicted by the $LHT$ model to the decay widths $\Gamma_{l\bar{l}}$
and $\Gamma_{inv}$ are calculated in sections 3 and 4, respectively.
In these sections, we also compare our numerical results with the
experimental measurement values and try to obtain constraints on the
free parameters of the $LHT$ model. Our conclusions are given in
section 5.

\noindent{\bf 2. The lepton flavor violation decay $Z \rightarrow
l\bar{l'}$}

The $LHT$ model [14] is based on an $SU(5)/SO(5)$ global symmetry
breaking pattern. A subgroup $[SU(2) \times U(1)]_{1} \times [SU(2)
\times U(1)]_{2}$ of the $SU(5)$ global symmetry is gauged, and at
the scale $f$ it is broken into the $SM$ electroweak symmetry
$SU(2)_{L}\times SU(1)_{Y}$. T-parity is an automorphism that
exchanges the $[SU(2) \times U(1)]_{1}$ and $[SU(2) \times
U(1)]_{2}$ gauge symmetries. The T-even combinations of the gauge
fields are the electroweak gauge bosons, and the T-odd combinations
are their T-parity partners. After taking into account electroweak
symmetry breaking, at the order of $\nu^{2}/f^{2}$, the
masses of the $T$-odd set of the $SU(2)\times U(1)$ gauge bosons are
given by
\begin{eqnarray}
 M_{B_{H}}=\frac{g'f}{\sqrt{5}}[1-\frac{5\nu^{2}}{8f^{2}}],\hspace{0.5cm}
 M_{Z_{H}}\approx
  M_{W_{H}}=gf[1-\frac{\nu^{2}}{8f^{2}}],
\end{eqnarray}
where $g'$ and $g$ are the $SM$ $U(1)_{Y}$ and $SU(2)_{L}$ gauge
coupling constants, respectively. $\nu=246GeV$ is the electroweak
scale.

To avoid severe constraints and simultaneously implement $T$ parity,
one needs to double the SM fermion doublet spectrum [14, 15]. The
$T$-even combination is associated with the SM $SU(2)_{L}$ doublet,
while the $T$-odd combination is its $T$-parity partner. The T-odd
fermion sector consists of three generations of mirror quarks and
leptons with vectorial couplings under $SU(2)_{L}\times U(1)_{Y}$.
Only T-odd leptons are related to our calculation, and we denote them by
\begin{equation}
\hspace{-1.5cm}
 \left(
 \begin{array}{c}\nu_{H}^{1}\\l_{H}^{1}\end{array}\right),
 \hspace{1.5cm} \left(
 \begin{array}{c}\nu_{H}^{2}\\l_{H}^{2}\end{array}\right),
 \hspace{1.5cm} \left(
 \begin{array}{c}\nu_{H}^{3}\\l_{H}^{3}\end{array}\right)
\end{equation}
with their masses satisfying to first order in $v/f$ [16]
\begin{equation}
\hspace{-1.5cm}
 M_{\nu_{H}}^{1}=M_{l_{H}}^{1}
 \hspace{1.5cm} M_{\nu_{H}}^{2}=M_{l_{H}}^{2},
 \hspace{1.5cm} M_{\nu_{H}}^{3}=M_{l_{H}}^{3}.
\end{equation}

The T-odd leptons (mirror leptons) have new flavor violating
interactions with the $SM$ leptons mediated by the T-odd gauge
bosons and at higher order by the triplet scalar $\Phi$, which are
parameterized by two $CKM$-like unitary mixing matrices $V_{Hl}$ and
$V_{H\nu}$. They satisfy $V_{H\nu}^{+}V_{Hl}=V_{PMNS}$, in which the
$PMNS$ matrix $V_{PMNS}$ is defined through neutrino mixing. As no
constraints on the $PMNS$ phases exist, we will set the three
Majorana phases of $V_{PMNS}$ to equal zero in our numerical
estimations, which is similar with Refs.[16,17].

\begin{figure}[htb]
\begin{center}
\epsfig{file=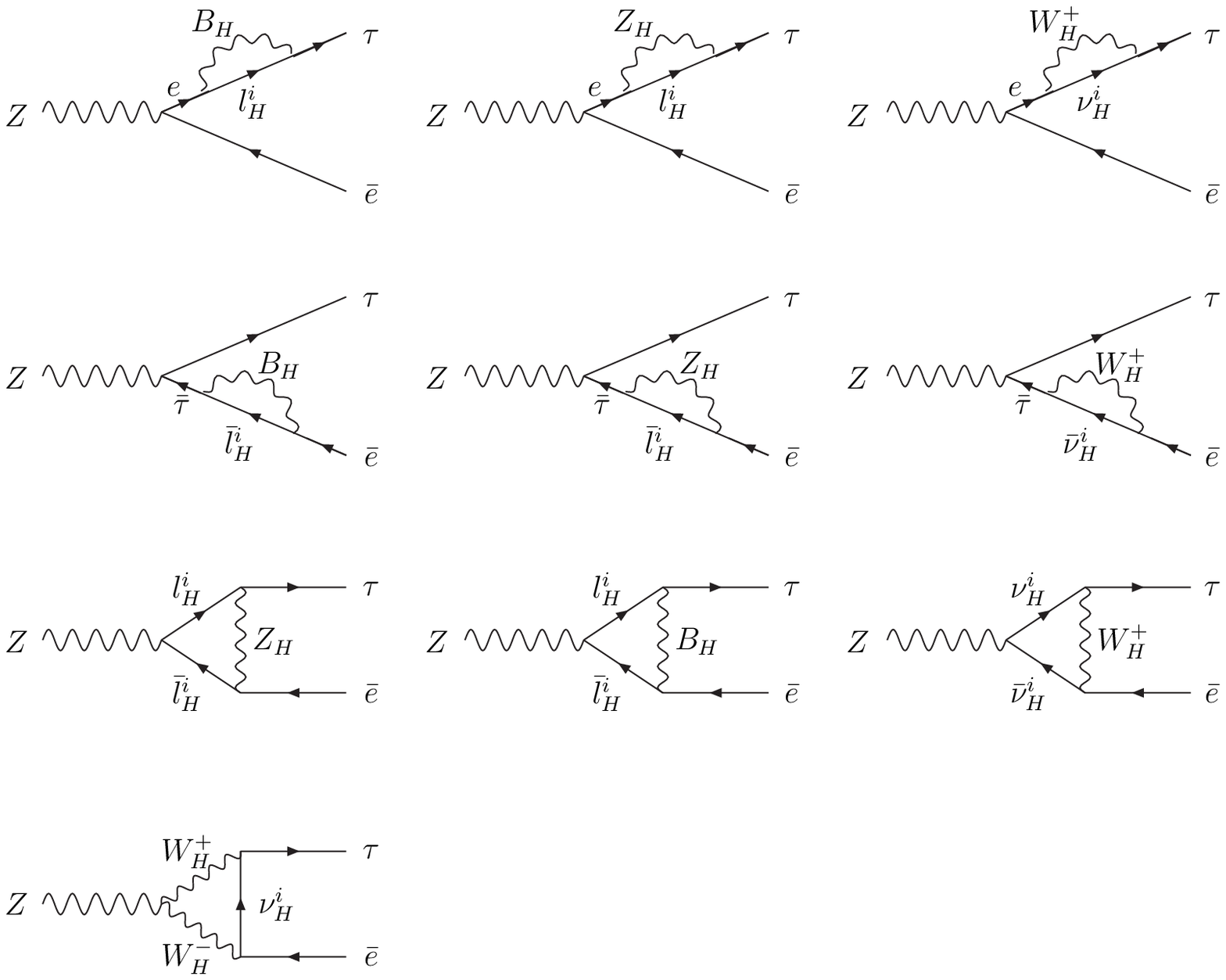,width=355pt,height=280pt} \caption{ The leading
Feynman diagrams for the $LFV$
 $Z$ decay $Z\rightarrow \tau \bar{e}$ in the $LHT$ \hspace*{ 2cm}model.}
 \label{ee}
\end{center}
\end{figure}

From the above discussions, we can see that the $LHT$ model provides
a new mechanism for lepton flavor violation, which comes from the
flavor mixing in the mirror lepton sector. It has been shown that
the $LHT$ model can give significant contributions to some $LFV$
processes, such as $l_{i}\rightarrow l_{j}\gamma$, $l_{i}\rightarrow
l_{j}l_{k}l_{l}$, $\tau\rightarrow \mu \pi$ $etc.$[16,18]. In the present
paper, we first consider the contributions of the $LHT$ model to the
$LFV$ $Z$ decay process $Z\rightarrow l\bar{l'}$. The relevant
Feynman diagrams for $Z\rightarrow \tau \bar{e}$ are shown in Fig.1.
The Feynman diagrams for the $LFV$ decay processes $Z\rightarrow
\tau \bar{\mu}$ and $Z\rightarrow \mu \bar{e}$ are similar to
Fig.1.

The $LHT$ model also predicts the existence of the T-odd scalar
triplet $\Phi$ with mass $M_{\Phi}$ of order $TeV$. Neglecting the
mass splitting between various components of the T-odd scalar
triplet $\Phi$, its contributions to the electroweak parameters $S$,
$T$, and $U$ vanish[19]. Ref.[19] has also shown that the effects of
$\Phi$ on the precision electroweak observables decouple with
growing $M_{\Phi}$. Furthermore, the T-odd scalar triplet $\Phi$ can
contribute to the $LFV$ $Z$ decay process $Z\rightarrow l\bar{l'}$
at the order higher than $v^{2}/f^{2}$ via its couplings to the
T-odd leptons and ordinary leptons. Thus, as a numerical estimation,
we will neglect its contributions in this paper, and the relevant
Feynman diagrams have not been shown in Fig.1.

\vspace*{2cm}
\begin{figure}[htb]
\begin{center}
\vspace{-0.5cm}
 \epsfig{file=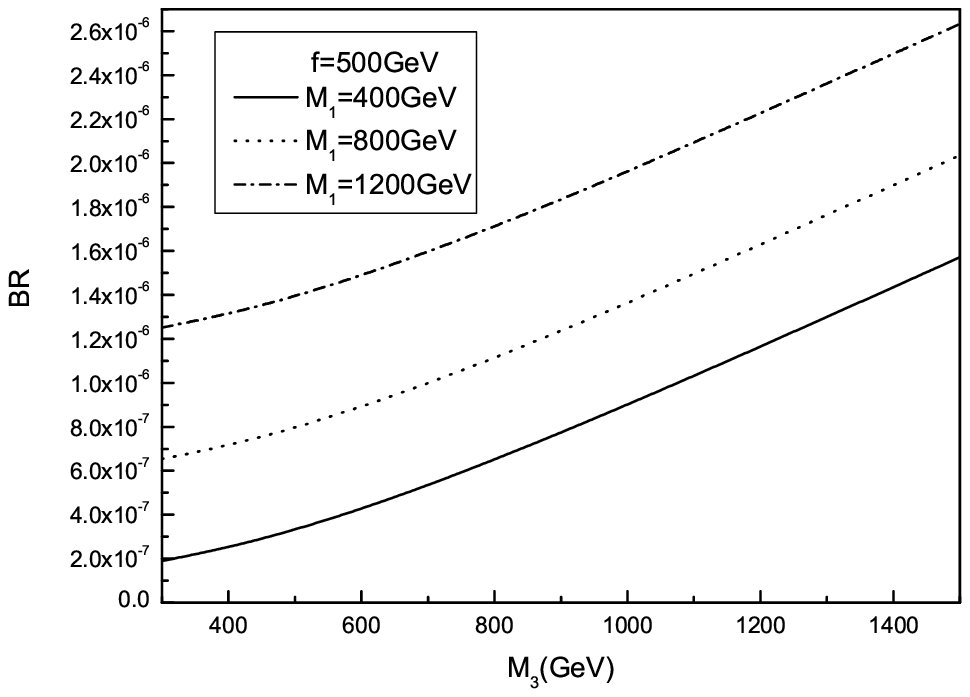,width=200pt,height=165pt}
\put(-115,-10){ (a)}\put(115,-10){ (b)}
 \hspace{0cm}\vspace{-0.25cm}
\epsfig{file=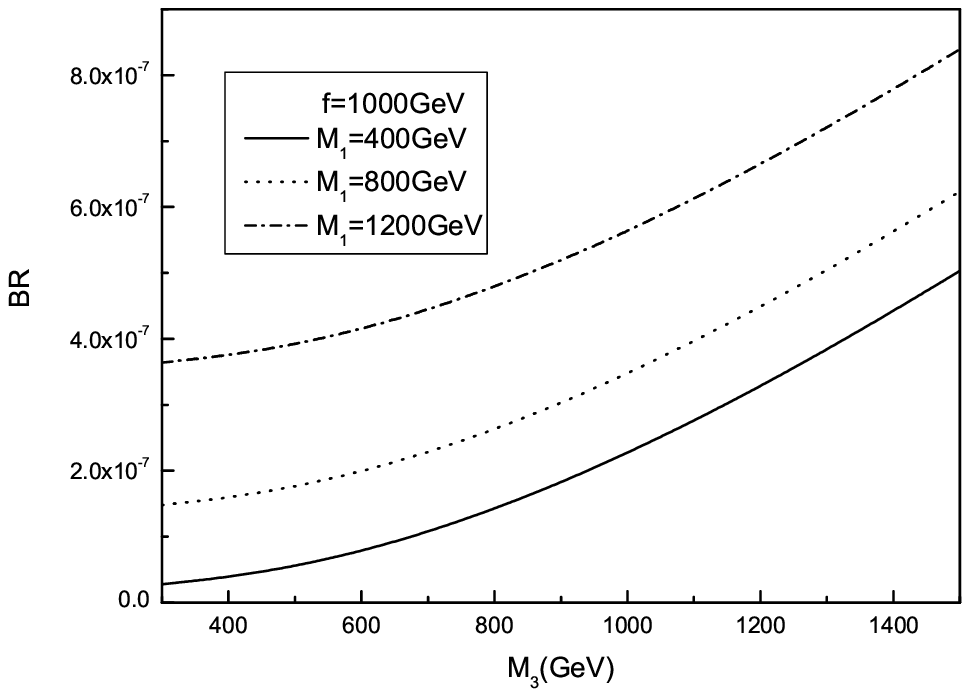,width=200pt,height=165pt} \hspace{-0.5cm}
 \hspace{10cm}\vspace{-1cm}
\vspace{0.5cm}
 \caption{The branching ratio $BR(Z\rightarrow \tau \bar{e})$ as a
 function of the third family T-odd \hspace*{2.0cm}lepton mass $M_{3}$ for
 $f=500$(a)and $1000GeV$(b). We have taken $M_{l_{H}}^{1}=M_{l_{H}}^{2}
 =\hspace*{2.0cm}M_{1}=400$,
 $800$, and $1200GeV.$}
 \label{ee}
\end{center}
\end{figure}
\begin{figure}[htb]
\begin{center}
\vspace{-0.5cm}
 \epsfig{file=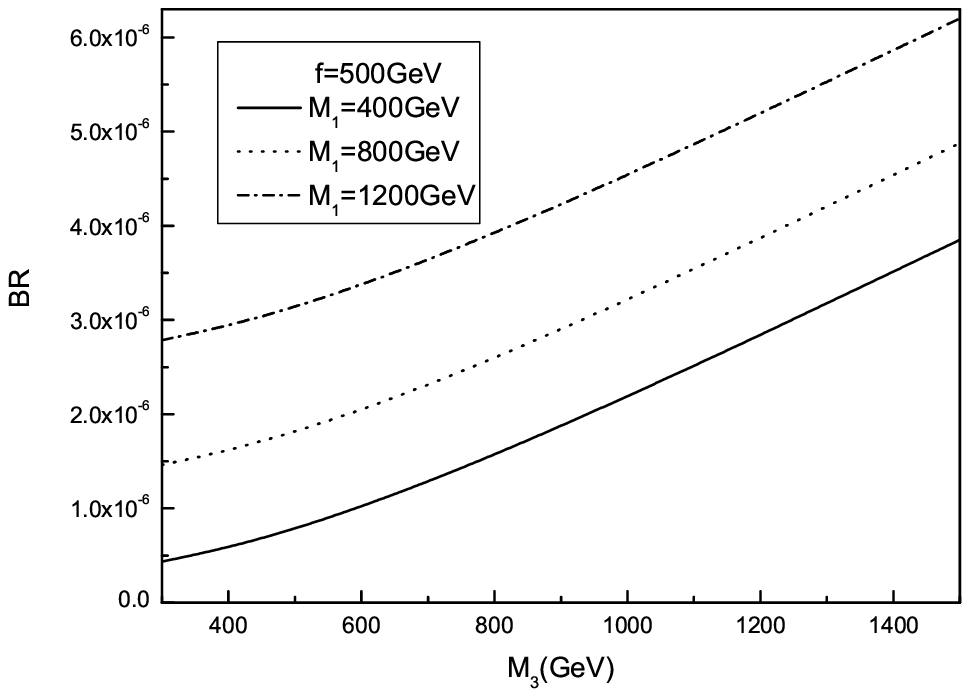,width=200pt,height=165pt}
\put(-115,-10){ (a)}\put(115,-10){ (b)}
 \hspace{0cm}\vspace{-0.25cm}
\epsfig{file=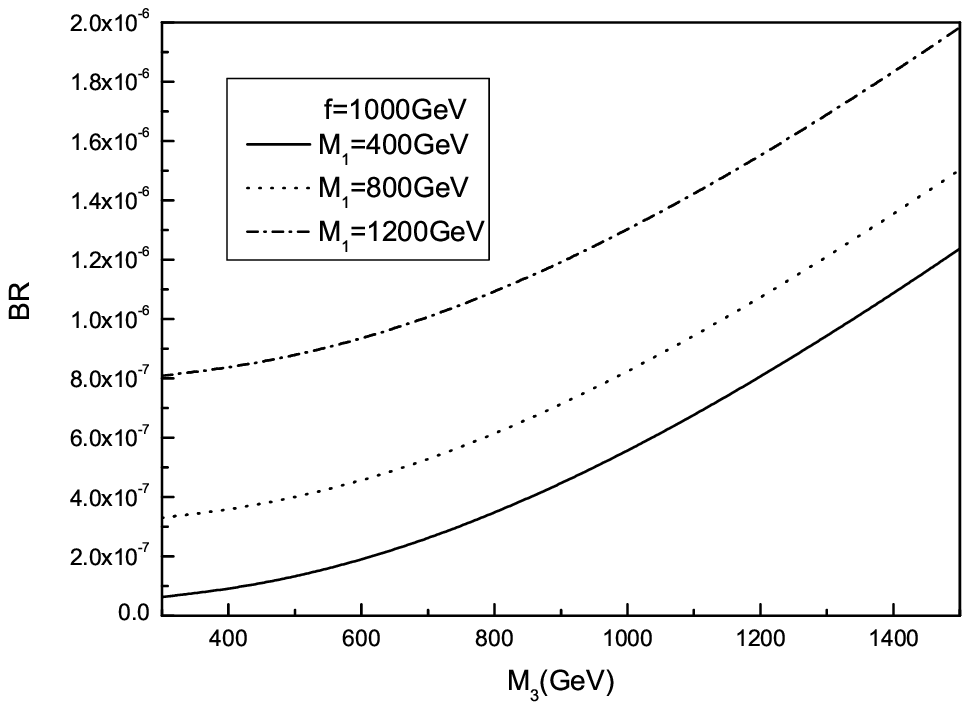,width=200pt,height=165pt} \hspace{-0.5cm}
 \hspace{10cm}\vspace{-1cm}
\vspace{0.5cm}
 \caption{The same as Fig.2 but for the $LFV$ decay process $Z\rightarrow
 \tau \bar{\mu}$.}
 \label{ee}
\end{center}
\end{figure}

\begin{figure}[htb]
\begin{center}
\vspace{-0.5cm}
 \epsfig{file=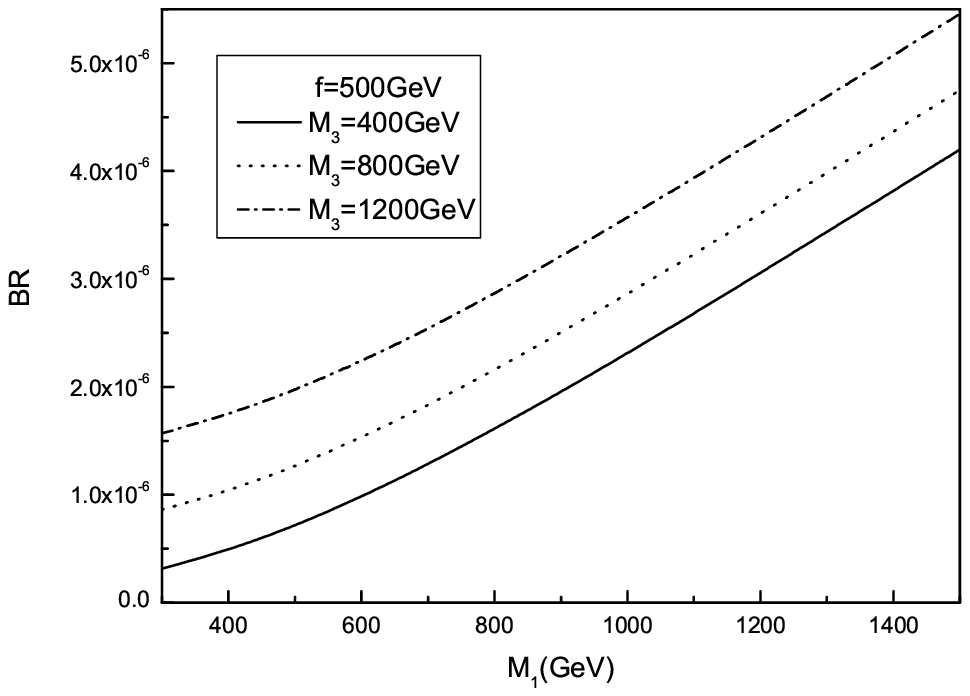,width=200pt,height=165pt}
\put(-115,-10){ (a)}\put(115,-10){ (b)}
 \hspace{0cm}\vspace{-0.25cm}
\epsfig{file=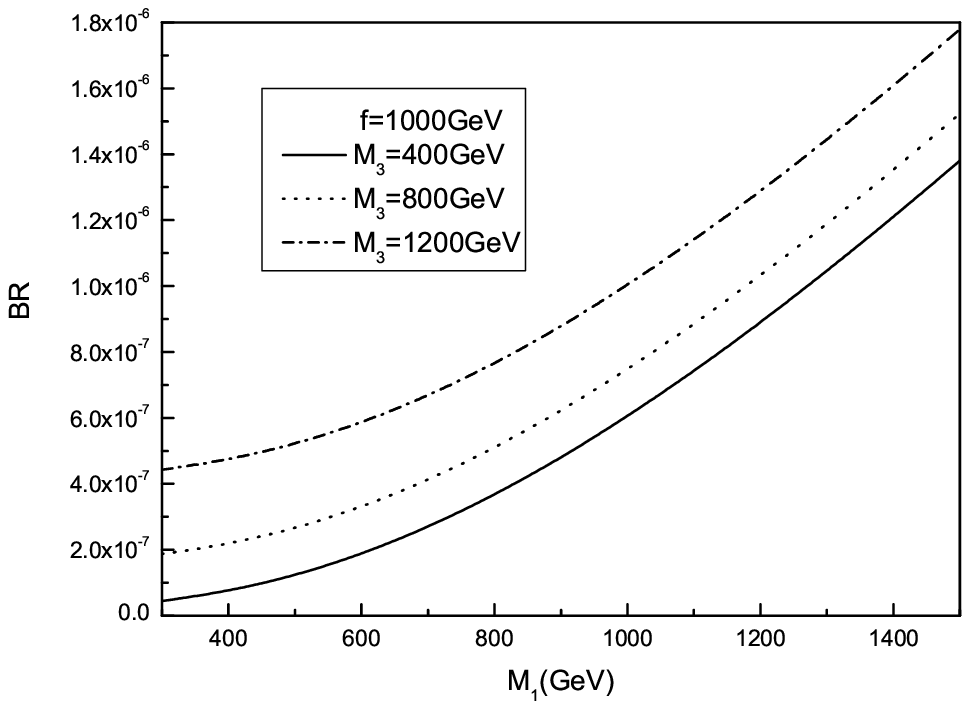,width=200pt,height=165pt} \hspace{-0.5cm}
 \hspace{10cm}\vspace{-1cm}
\vspace{0.5cm}
 \caption{The branching ratio $BR(Z\rightarrow \mu\bar{e})$ as a function of the
 mass parameter $M_{1}$ \hspace*{2.0cm}$for f=500$(a) and $1000GeV$(b). We have
  taken $M_{l_{H}}^{1}= M_{l_{H}}^{2}=M_{1}$ and \hspace*{2.0cm} $ M_{3}=400,
  800, 1200GeV$}.
 \label{ee}
\end{center}
\end{figure}
The amplitude of the $LFV$ decay $Z\rightarrow l\bar{l'}$ is given
by
\begin{eqnarray}
 M(Z\rightarrow l\bar{l'})=\varepsilon^{\mu}\bar{u}(p)\Gamma_{\mu}u(p'),
\end{eqnarray}
where $p$ and $p'$ are the momenta of the leptons $l$ and $l'$,
respectively. $\varepsilon^{\mu}$ is the polarization vector of the
on-shell gauge boson $Z$. The effective vertex $\Gamma_{\mu}$ can be
obtained via calculating Fig.1, which can be generally written as:
\begin{eqnarray}
 \Gamma_{\mu}=\gamma_{\mu}(f_{V}-f_{A}\gamma_{5})+q^{\nu}(if_{M}+f_{E}\gamma_{5})
 \sigma_{\mu\nu},
\end{eqnarray}
where $\sigma_{\mu\nu}=\frac{i}{2}[\gamma_{\mu},\gamma_{\nu}]$ and
$q$ is the momentum transfer with $q^{2}=(p-p')^{2}$. The form
factors $f_{V}$, $f_{A}$, $f_{M}$ and $f_{E}$ include all of the
contributions from the diagrams in Fig.1. For simplicity, we omit
the explicit expressions for these form factors. In calculations of
the one-loop diagrams, we have used LOOPTOOLS [20] and ignored the
masses of the final state leptons $l$ and $l'$.

It is obvious that, except for the $SM$ input parameters
$\alpha_{e}=1/128.8$, $S_{W}^{2}=0.2315$, and $M_{Z}=91.187GeV$ [2],
the branching ratio $BR(Z\rightarrow l\bar{l'})$ is dependent on the
model--dependent parameters $f$, $(V_{Hl})_{ij}$, and the T-odd
leptons masses . The matrix elements $(V_{Hl})_{ij}$ can be
determined through $V_{Hl}=V_{H\nu}V_{PMNS}$. To avoid any
additional parameters introduced and to simply our calculations, we
take $V_{Hl}=V_{PMNS}$ and $V_{H\nu}=I$, which means that the T-odd
leptons have no effects on the flavor violating observables in the
neutrino sector [16,18]. For the $PMNS$ matrix $V_{PMNS}$, we take
the standard parameterization form with parameters given by the
neutrino experiments [21].

Our numerical results are summarized in Fig.2, Fig.3 and Fig.4, in
which we have plotted the $BRs$ as functions of the T-odd lepton
mass for $f=500$ and $1000GeV$. For Figs.2 and 3, which
correspond the $LFV$ processes $Z\rightarrow \tau \overline{e}$ and
$Z\rightarrow \tau \overline{\mu}$, respectively, we have taken
$M_{l_{H}}^{1}=M_{l_{H}}^{2}=M_{1}$ and $M_{l_{H}}^{3}=M_{3}$. For
the $LFV$ process $Z\rightarrow \mu \overline{e}$ given in Fig.4, we
have taken $M_{l_{H}}^{2}=M_{l_{H}}^{3}=M_{3}$ and
$M_{l_{H}}^{1}=M_{1}$. One can see from these figures that the
contributions of the $LHT$ model to the $LFV$ process $Z\rightarrow
l\bar{l'}$ increase as the T-odd lepton mass increases and the
scale parameter $f$ decreases. In most of the parameter space, the
values of the branching ratios $BR(Z\rightarrow \tau \overline{e})$
and $BR(Z\rightarrow \tau \overline{\mu})$ can not overcome the
current experimental limits given in Eq.(5), while they can overcome the
improved sensitivities at $Giga-Z$, given in Eq.(7). For the $LFV$
process $Z\rightarrow \mu \bar{e}$, its current experimental limit
can give severe constraints on the free parameters of the $LHT$
model. If one would like to reduce the contributions of the $LHT$
model to the $LFV$ process $Z\rightarrow l\bar{l'}$ and make its
$BR$ value satisfy the current or future experimental limits, one
has to enhance the value of the scale parameter $f$, reduce the mass
splitting between three generations of the T-odd leptons, or make
the matrix $V_{Hl}$ much more hierarchical than the $PMNS$ matrix
$V_{PMNS}$.

\noindent{\bf 3. The lepton flavor conservation decay $Z \rightarrow
l\bar{l}$}

In the $SM$, the $LFC$ decay process $Z
\rightarrow l\bar{l}$ exists at the tree level. The partial $Z$
decay width $\Gamma(Z\rightarrow l\bar{l})$ including QED and QCD
corrections can be written as [2]
\begin{eqnarray}
\Gamma(Z\rightarrow
l\bar{l})=\frac{G_{F}M_{Z}^{3}}{6\sqrt{2}\pi}[(\bar{g}^{l}_{V})^{2}
+(\bar{g}^{l}_{A})^{2}](1+\delta\rho+\delta\rho_{l}+\delta_{QED}).
\end{eqnarray}
The vector and axial-vector $Zl\bar{l}$ couplings $\bar{g}^{l}_{V}$
and $\bar{g}^{l}_{A}$ comprise one-loop and higher electroweak and
internal QCD corrections through the form factors $\delta\rho_{l}$
and $k_{l}$, which can be written as
\begin{eqnarray}
\bar{g}^{l}_{V}=\sqrt{\rho_{l}}(\frac{1}{2}-2\sin^{2}\theta_{eff}^{l}),\hspace*{0.5cm}
\bar{g}^{l}_{A}=\sqrt{\rho_{l}}\times\frac{1}{2},
\end{eqnarray}
with $\sin^{2}\theta_{eff}^{l}=k_{l}\sin^{2}\theta_{W}$, in which
$\theta_{W}$ is the Weinberg angle. The term $\delta\rho$ is the
deviation from the SM prediction for the $\rho$ parameter
$\rho=M_{Z}\cos\theta_{W}/M_{W}=1+\delta\rho$, and $\delta_{QED}$
accounts for the final state photon radiation.
\begin{figure}[htb]
\begin{center}
\epsfig{file=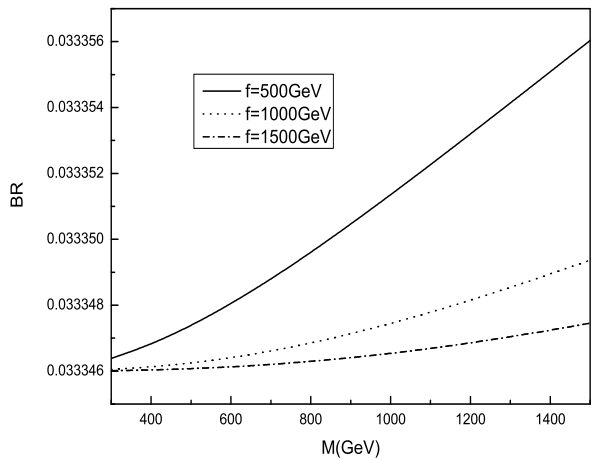,width=245pt,height=200pt}
 \vspace{-0.4cm}\caption{The branching ratio $BR(Z\rightarrow e\bar{e})$ as
 a function of the
 mass parameter $M$ for  \hspace*{1.8cm}three values of the parameter
 $f$. We have assumed $M_{l_{H}}^{1}=
 M_{l_{H}}^{2}=M_{l_{H}}^{3}=M$.}
\end{center}
\end{figure}
\begin{figure}[htb]
\begin{center}
\vspace{-0.5cm}
 \epsfig{file=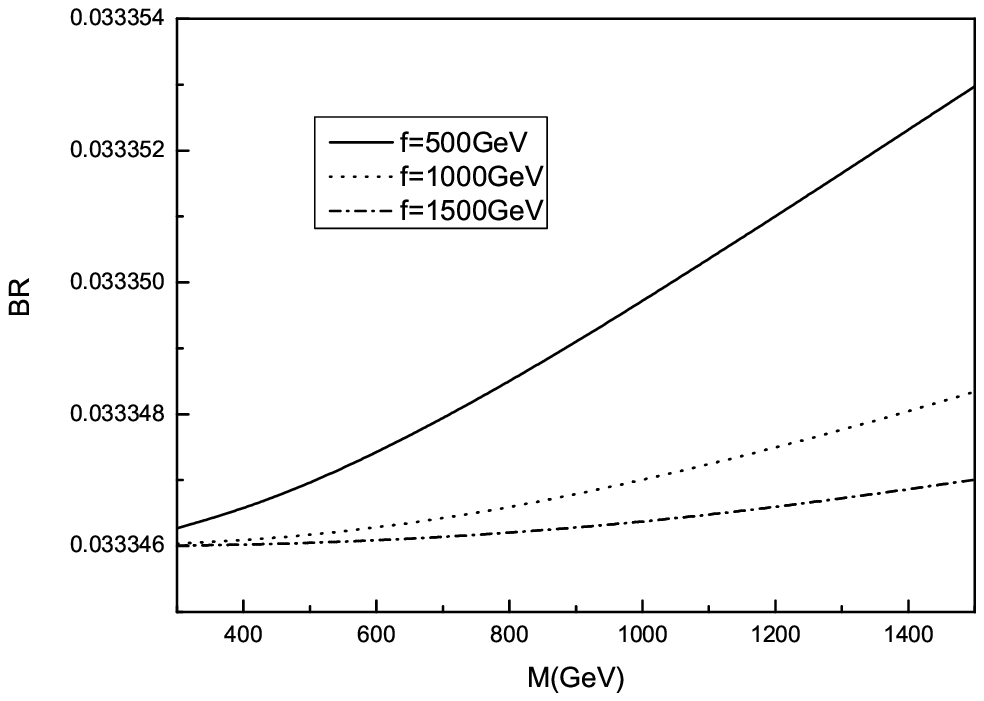,width=200pt,height=165pt}
\put(-115,-10){ (a)}\put(115,-10){ (b)}
 \hspace{0cm}\vspace{-0.25cm}
\epsfig{file=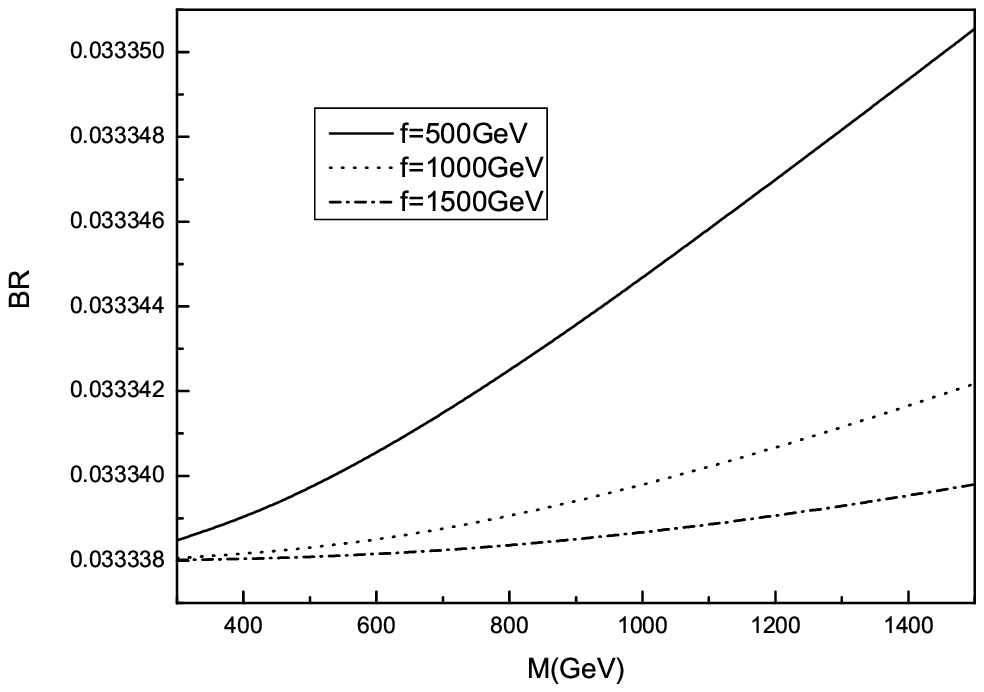,width=200pt,height=165pt} \hspace{-0.5cm}
 \hspace{10cm}\vspace{-1cm}
\vspace{0.5cm}
 \caption{The same as Fig.5 but for the $LFC$ decay processes $Z\rightarrow
 \mu\bar{\mu}$ (a) and $Z\rightarrow \tau\bar{\tau}$ (b).}
 \label{ee}
\end{center}
\end{figure}

 In the $LHT$ model, all of the $SM$ particles are assigned with an
even T-parity, while all of the new particles are assigned with an odd
T-parity, except for the little $Higgs$ partner of the top quark. If
the T-parity is an exact symmetry, the $SM$ gauge bosons do not mix
with the T-odd gauge bosons, and thus the electroweak observables
are not modified at the tree level. So the $LHT$ model can only
contribute the partial width $\Gamma(Z\rightarrow l \bar{l})$ at the
one-loop level. According to discussions given in section 2, the
one-loop contributions of the T-odd triplet scalar $\Phi$ to the
decay width $\Gamma(Z\rightarrow l\bar{l})$ can be neglected. The
contributions of the $LHT$ model to the Weinberg angle $\theta_{W}$
and the parameter $\rho$ have been extensively studied in
Refs.[14,19]. It has been shown that, as long as the scale parameter
$f\geq$500GeV, the $LHT$ model can be consistent with precision
electroweak data. However, the T-odd leptons have contributions to
$\Gamma(Z\rightarrow l\bar{l})$ via correcting the parameter
$\rho_{l}$. The relevant Feynman diagrams are similar with those
given in Fig.1, only assuming $l'=l=e,\mu,$ or $\tau$.

In this section, we focus our attention on the contributions of the
$LHT$ model to the LFC decay process $\Gamma(Z\rightarrow l
\bar{l})$. So in our numerical estimation, we will take the T-odd
leptons degenerating in mass and assume $M_{l_{H}}=M_{\nu _{H}}=M$.
This means that the T-odd leptons have no contributions to the $LFV$
processes, which is the minimal flavor violation ($MFV$) limit of
the $LHT$ model [18,22]. In this case, the decay width
$\Gamma(Z\rightarrow l\bar{l})$ depends on the mass parameter $M$,
the scale parameter $f$, and the unitary mixing matrix $V_{Hl}$.
Similar with section 2, we also take $V_{Hl}=V_{PMNS}$. Our
numerical results are given in Figs.5 and 6, in which we plot the
branching ratio $BR(Z\rightarrow l \bar{l})$ ($l=e, \mu,$ or $\tau$)
as a function of the mass parameter $M$ for three values of the
scale parameter $f$. One can see from these figures that the $LHT$
model generates the positive contributions to these branching ratios
$BR(Z\rightarrow e\bar{e})$, $BR(Z\rightarrow \mu\bar{\mu})$, and
$BR(Z\rightarrow \tau\bar{\tau})$. Their values increase as $M$
increases and $f$ decreases. However, in most of the parameter
space of the $LHT$ model, the values of the $BRs$ cannot overcome
their experimental measurement values.

\noindent{\bf 4. The invisible $Z$ decay $Z \rightarrow
\nu\bar{\nu}$}

\begin{figure}[htb]
\begin{center}
\epsfig{file=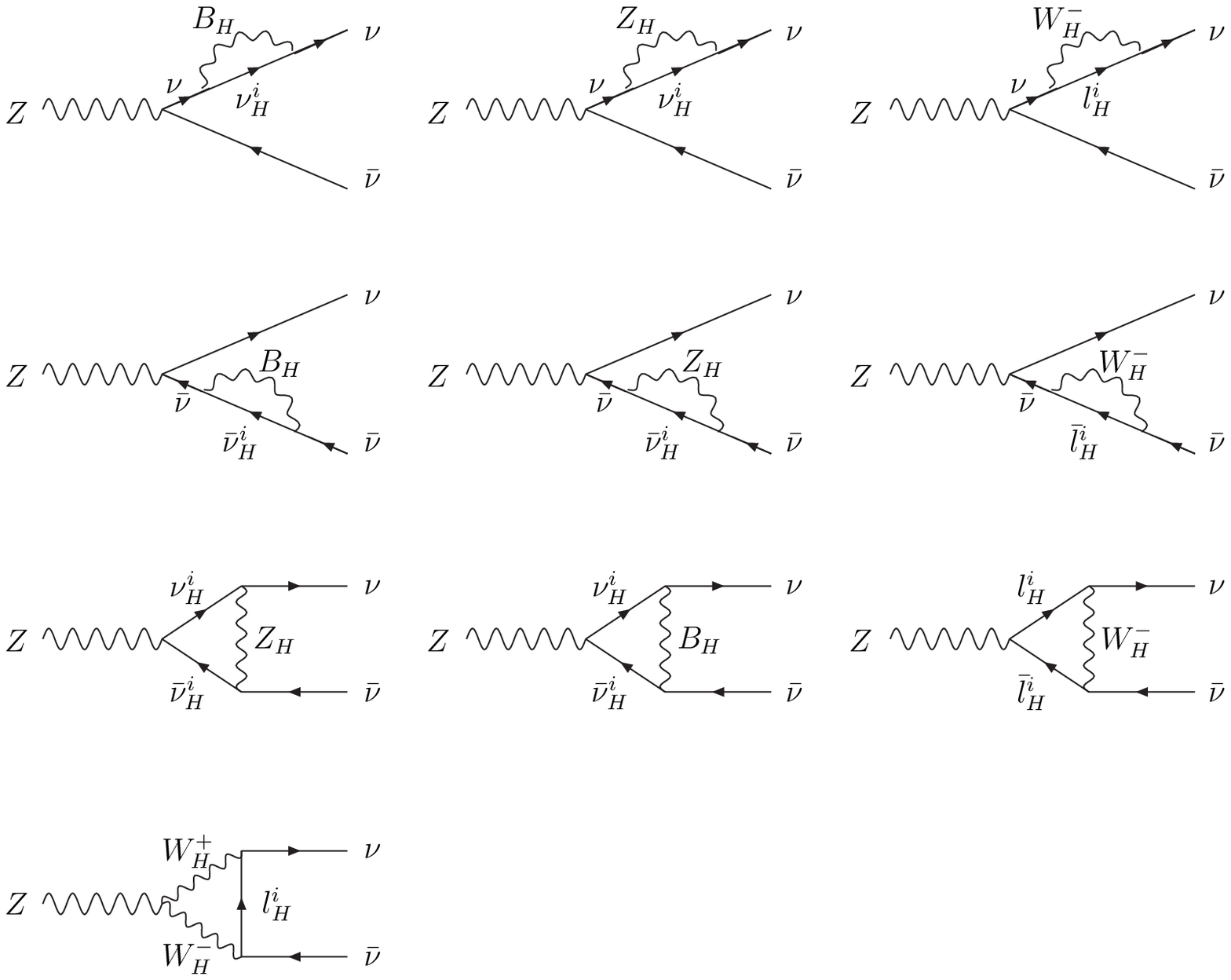,width=395pt,height=530pt} \vspace{-7.3cm}
\caption{ The leading Feynman diagrams for the $LFV$ $Z$ decay
$Z\rightarrow \nu
 \bar{\nu}$ in the $LHT$ \hspace*{2cm} model.}
 \label{ee}
\end{center}
\end{figure}
The $SM$ has been extensively tested by experiments at the $CERN$
$e^{+}e^{-}$ collider $LEP$, the Fermilab Tevatron, and elsewhere. At the
$LEP$, the coupling of the gauge boson $Z$ to neutrinos is
constrained by the invisible $Z$ decay width $\Gamma_{inv}$, which
receives contributions from all neutrinos flavors. Thus, it is
possible to constrain new physics contributions to the
$Z\nu\bar{\nu}$ coupling that respect universality.

\begin{figure}[htb]
\begin{center}
\epsfig{file=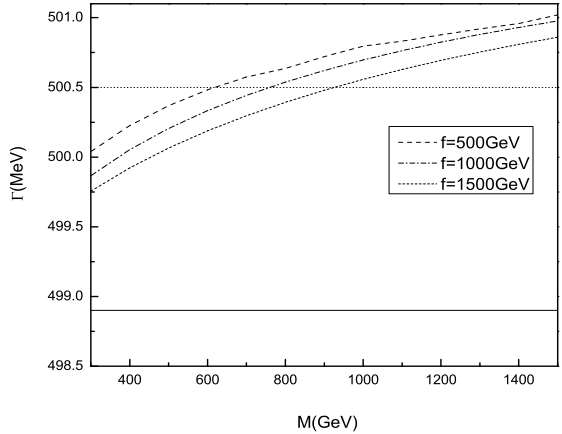,width=245pt,height=200pt}
 \vspace{-0.4cm}\caption{The invisible $Z$ decay width $\Gamma_{inv}$ as a
 function of $M$ for three values of $f$. The \hspace*{1.92cm}horizontal
 solid and dotted lines denote the central and
 upper values of $\Gamma^{exp}_{inv}$,
  \hspace*{1.92cm}respectively.}
\end{center}
\end{figure}

From the above discussions, we can see that the $LHT$ model can
contribute the $Z\nu\bar{\nu}$ coupling at the one-loop level. The
relevant Feynman diagrams are plotted in Fig.7. Similar to Fig.1,
we also neglect the contributions of the T-odd triplet scalar
$\Phi$. The contributions of the $LHT$ model to the invisible $Z$
decay width $\Gamma_{inv}$ are dependent on the mass parameter
$M_{\nu_{H}}=M_{l_{H}}=M$, the unitary mixing matrix $V_{H\nu}$, and
the scale parameter $f$. In this section, we also assume
$V_{H\nu}=I$. In Fig.8, we plot the invisible $Z$ decay width
$\Gamma_{inv}$ including the contributions of the $LHT$ model as a
function of the mass parameter $M$ for three values of the parameter
$f$. To compare our calculation value with the experimental value
$\Gamma^{exp}_{inv}=499\pm1.5GeV$ and see whether it can give new
constraints on the $LHT$ model, we give $\Gamma^{exp}_{inv}$ in
Fig.8, in which the horizontal solid  and dotted lines indicate the
 central and maximal values of the experimental measurement for
 $\Gamma_{inv}$, respectively. One can see from
Fig.8 that, in the case of $V_{H\nu}=I$ and
$M_{\nu_{H}}=M_{l_{H}}=M$, if one demands the $LHT$ prediction value
for $\Gamma_{inv}$ to be in the ranges allowed by the $LEP$
experiments, the mass parameter $M$ must be smaller than $700GeV$
for $f\leq1000GeV$ and smaller than $900GeV$ for $f\leq1.5TeV$.

\noindent{\bf 5. Conclusions}

In order to implement T-parity in the fermion sector of the $LHT$
model, the T-odd $SU(2)$ doublet fermions, which are called the
mirror fermions of the $SM$ fermions, have to be introduced. The
mirror fermions can couple to ordinary fermions mediated by the
T-odd gauge bosons and at higher order by the T-odd scalar triplet
$\Phi$. Thus, these new fermions can generate correction effects on
some observables at the one-loop level. Furthermore, flavor mixing
in the mirror fermion sector gives rise to a new source of flavor
violation, which might generate significant contributions to some
flavor violation processes.

The $SM$ gauge boson $Z$ will be abundantly produced at the  $LHC$
and the future high energy linear $e^{+}e^{-}$ collider experiments.
It is possible to examine its properties with unprecedented
precision. In this paper, we consider the contributions of the $LHT$
model to the leptonic $Z$ decays. Our numerical results show that if
one demands the branching ratio $BR(Z\rightarrow l\bar{l'})$ below
the present (for $Z\rightarrow \mu \bar{e}$) and the future (for
$Z\rightarrow \tau \bar{\mu}$ and $Z\rightarrow \tau \bar{e}$)
experimental upper bounds, the relevant mixing matrix $V_{Hl}$ must
be rather hierarchical, unless the spectrum of the T-odd leptons is
quasidegenerate. Our conclusions are similar with those given by
Refs.[16,18]. For the $LFC$ decay $Z\rightarrow l\bar{l}$, the $LHT$
model can give positive contributions, which is favored by the
current high energy collider experiments. However, the current
experimental values for $BR(Z\rightarrow l\bar{l})$ ($l=\tau$,
$\mu$, and  $e$) can not give severe constraints on the free
parameters of the $LHT$ model, although the coupling of the gauge
boson $Z$ to individual neutrino flavor has not been tested with
comparably good accuracy. The couplings of the gauge boson $Z$ to
three family neutrino flavors can be constrained by the measurement
invisible $Z$ decay width $\Gamma_{inv}$. In this paper we also
calculate the contributions of the $LHT$ model to the invisible $Z$
decay width and compare our result with its experimental value. We
find that the upper limit of $\Gamma_{inv}^{exp}$ can give
constraints on the free parameters $M$ and $f$.

\vspace{2.0cm}

\noindent{\bf Acknowledgments}

This work was supported in part by the National Natural Science
Foundation of China under Grants No.10675057, the Natural Science
Foundation of the Liaoning Scientific Committee(No.20082148) and
Foundation of Liaoning  Educational Committee(No.2007T086).

\end{document}